\documentclass[11pt,a4paper]{article}
\usepackage{jcappub}
\usepackage{mathrsfs}
\usepackage{amsmath,amssymb,graphics,epsfig,subfigure}
\usepackage{color}

\title{Constraints on running vacuum models with the baryon-to-photon ratio}
\author{Hao Yu$~^{a,}$\footnote{yuhaocd@cqu.edu.cn},
Ke Yang~$^{b,}$\footnote{keyang@swu.edu.cn},
Jin Li~$^{a,}$\footnote{cqujinli1983@cqu.edu.cn, corresponding author}}
\affiliation{
$^a$ Physics Department, Chongqing University, Chongqing 401331, China\\
$^b$ School of Physical Science and Technology, Southwest University, Chongqing 400715, China}

\abstract
{We study the influence of running vacuum on the baryon-to-photon ratio in running vacuum models (RVMs). When there exists a non-minimal coupling between photons and other matter in the expanding universe, the energy-momentum tensor of photons is no longer conserved, but the energy of photons could remain conserved. We discuss the conditions for the energy conservation of photons in RVMs. The photon number density and baryon number density, from the epoch of photon decoupling to the present day, are obtained in the context of RVMs by assuming that photons and baryons can be coupled to running vacuum, respectively. Both cases lead to a time-evolving baryon-to-photon ratio. However the evolution of the baryon-to-photon ratio is strictly constrained by observations. It is found that if the dynamic term of running vacuum is indeed coupled to photons or baryons, the coefficient of the dynamic term must be extremely small, which is unnatural. Therefore, our study basically rules out the possibility that running vacuum is coupled to photons or baryons in RVMs.}


\begin{document}


\maketitle

\section{Introduction}
The Big Bang theory is by far the most accepted theory explaining the birth, evolution and future of the universe. The measurements of the hydrogen abundance and helium abundance in the universe are exactly consistent with the predictions given by the standard Big Bang nucleosynthesis (BBN) model, which is regarded as one of the strongest evidences\footnote{The other two evidences are the cosmic microwave background radiation and observed expansion of the universe.} to support the Big Bang theory (see Refs.~\cite{Malaney:1993ah,Sarkar:1995dd,
Tytler:2000qf,Fields:2014uja,Mathews:2017xht,Zyla:2020zbs} for a review). One of the ingenuities of the Big Bang theory is that it proposes that the abundance of light elements in the universe is only related to one basic parameter, the baryon-to-photon ratio $\eta$. Therefore, the photon and baryon number densities are of interest and significant for cosmology. In the standard cosmological model (i.e., the $\Lambda$CDM model), they dilute, as the universe expands, at the same dilution rate after BBN, so the present-day value of the baryon-to-photon ratio is a constant and was formed at the end of BBN~\cite{Boesgaard:1985km,Steigman:2007xt,Cyburt:2015mya,
Mathews:2017xht}.

In different epochs of the universe, $\eta$ can be determined independently from different perspectives. For the BBN epoch one can derive $\eta$ based on the observational data on the abundance of primordial chemical elements~\cite{Boesgaard:1985km,Copi:1994ev,Sarkar:2002er,
Steigman:2007xt,Cyburt:2015mya,Mathews:2017xht,Zyla:2020zbs}, and for the primordial recombination epoch one can get $\eta$ by analysing the anisotropy of the cosmic microwave background (CMB) radiation~\cite{Bond:1993fb,Hu:1994jd,Jungman:1995bz,Hu:2001bc}. Whether these independent $\eta$'s evolving with the universe are in line with the current observed $\eta_0$~\cite{Fukugita:2004ee,Tanabashi:2018oca,Aghanim:2018eyx,
Fields:2019pfx,Zyla:2020zbs}, is the pivotal to judging the viability of the hypotheses and calculations in the $\Lambda$CDM model. In practice, the photon and baryon number densities are affected by many factors related to the cosmological model, especially the modified gravity theory and the components of the universe under consideration. If dark matter could annihilate or decay (see Refs.~\cite{Ullio:2002pj,Chen:2003gz,Padmanabhan:2005es,
Ando:2005xg,Bertone:2007aw,Galli:2009zc,Galli:2011rz} and references therein), then the baryon-to-photon ratio will deviate from the constant predicted in the $\Lambda$CDM model. But from BBN to the present, the deviation is almost negligible ($\frac{\Delta \eta}{\eta}\leq10^{-5}$)~\cite{Zavarygin:2015ave,Zavarygin:2015rba}. The impact of dark energy on $\eta$ is mainly due to the possibility that dark energy could decay into CMB, but the decay rate of dark energy must be sufficiently small to be in accordance with observations~\cite{Peebles:2002gy,Opher:2004vg,Opher:2005px,
Kumar:2018yhh,Yadav:2019jio}. Considering low-amplitude and large-scale fluctuations, the evolution of $\eta$ may change with the region of the universe~\cite{Barrow:2018yyg}. In addition, $\eta$ is influenced by varying fundamental ``constants'' as well~\cite{Campbell:1994bf,Dent:2001ga,Copi:2003xd,
Li:2005aia,Alvey:2019ctk}. The variation of Newton's constant could modify nucleosynthesis codes, which can be used for restraining Brans-Dicke theory and scalar-tensor theories~\cite{Clifton:2005xr,Alvey:2019ctk}. A time-varying cosmological ``constant'' is also able to affect the primordial nucleosynthesis~\cite{Freese:1986dd,Sarkar:1995dd,Birkel:1996py}.
Although most of these works do not specify the extent of the influence of varying fundamental ``constants'' on $\eta$, it is not hard to infer that, in these cosmological models, either $\eta$ is a constant different from the one in the $\Lambda$CDM model, or it will inevitably evolve over time. Overall, the prediction of $\eta$ is a powerful tool to verify the feasibility of various cosmological models.

In this work, we study the effect of varying vacuum on $\eta$ in a class of cosmological models, which are so called running vacuum models (RVMs), see Refs.~\cite{Sola:2011qr,Sola:2013gha,Sola:2015rra,Perico:2013mna} and references therein. There is a long history of investigations of dynamic vacuum~\cite{Terazawa:1981ga,Alvarenga:1998wx,Zhang:2006qu,
Stadnik:2014tta,Gomez-Valent:2014rxa,Li:2015vla}, but most dynamic vacuum models lack theoretical support and motivation. The proposal of RVMs is mainly attempting to connect the vacuum energy density with the quantum field theory in curved space-time~\cite{Shapiro:2000dz,Shapiro:2009dh,Sola:2007sv}. In recent years, the research and development of RVMs have shown that running vacuum is better than a single cosmological constant in some respects, such as describing cosmology global picture~\cite{Sola:2015rra,Lima:2012mu,Perico:2013mna,Geng:2020upf}, matching observational data~\cite{Gomez-Valent:2014rxa,Sola:2017znb,Perico:2016kbu,
Rezaei:2019xwo}, and alleviating tensions in cosmology~\cite{Sola:2016zeg,Geng:2017apd,Gomez-Valent:2017idt,
Sola:2017znb,Gomez-Valent:2018nib}. Recently, it is found that RVMs, comparing with the $\Lambda$CDM model, even exhibit more superior thermodynamic characteristics~\cite{Lima:2015mca,Gonzalez-Espinoza:2019vcy,
Sola:2019uum}.

In RVMs, vacuum is actually non-minimally coupled to matter fields. According to the regular definition of the energy-momentum tensor, it can be found that $\nabla_{\mu}T_m^{\nu\mu}\neq0$, where $T_m^{\mu\nu}$ is the energy-momentum tensor of all matter fields in RVMs. According to the viewpoint of Tiberiu Harko et al.~\cite{Harko:2014pqa,Harko:2015pma}, if the energy-momentum tensor of a matter field is not conserved, the field equation can be in comparison with the first law of thermodynamics for open systems. Therefore, $\nabla_{\mu}T_m^{\nu\mu}\neq0$, at the macroscopic level, can be explained as the production (or annihilation) of the corresponding particles. If photons in RVMs are coupled to vacuum, the redshift of a single photon in the expanding universe will be modified comparing with its ``free expansion'' in the $\Lambda$CDM model. We wonder whether such expansion of photons in RVMs could solve the problem of energy non-conservation in the $\Lambda$CDM model\footnote{Since the evolution of CMB radiation satisfies $\nabla_{\mu}T_r^{\nu\mu}=0$ in the $\Lambda$CDM model (it is also known as black-body radiation, see e.g.~\cite{Tsallis:1995zza,Lima:1996kc} and references therein), the total energy of CMB decreases with the expansion of the universe even though photons have no interaction with other matter fields. From this point of view, we have to define subtly, with only one tiny cosmological ``constant", the energy of space-time to guarantee energy conservation of the entire universe.}~\cite{Novello:1989zb,Macleod:2004rd,
Macleod:2005pi}. Furthermore, the impact of running vacuum on CMB results in a varying baryon-to-photon ratio, which must be consistent with current observations. Similarly, if running vacuum is coupled to baryons and the decay (or absorption) of vacuum is a baryon non-conserving process, there could be baryon production (or annihilation) even after the Hadron epoch. Since we have no any experiment and observation implying that the baryon non-conserving process can occur in the context of low energy and weak gravitational field, in order not to contradict the existing experiments and observations, the occurrence rate of running vacuum decaying into baryons (or the contrary process) must be very low. Therefore, it imposes strict constraints on the dynamic terms in the energy density of running vacuum.

It is worth noting that the influence of running vacuum on the baryon-to-photon ratio actually includes two aspects: indirect influence and direct influence. A typical case of the former is that evolving vacuum has more or less effect on the primordial nucleosynthesis, so it indirectly alters the baryon-to-photon ratio during the BBN epoch~\cite{Birkel:1996py}. The latter is that the coupling between dynamic vacuum and photons (or baryons) can directly affect the number of photons (or baryons) per comoving volume via particle production, so that the baryon-to-photon ratio evolves even after photon decoupling. In this work we only consider the latter case. More relevant literature on the indirect influence of dynamic ``constants'' on the baryon-to-photon ratio can be found in Refs.~\cite{Birkel:1996py,Nakamura:2005qm,Martins:2020syb}.

Finally, according to the current baryon-to-photon ratio $\eta_0$~\cite{Tanabashi:2018oca,Aghanim:2018eyx,Fields:2019pfx}, theoretically one can derive the value of $\eta$ at any epoch through assuming that vacuum is coupled to photons or baryons. However, before photon decoupling (which is closely related to the end of recombination), photons are not only coupled to running vacuum, but also to baryon nuclei, which makes our research about the impact of running vacuum on the baryon-to-photon ratio complicated. So, in this work we only discuss the evolution of the universe after photon decoupling. By estimating the value of $\eta$ at the epoch of photon decoupling (based on the assumption that running vacuum is non-minimally coupled to photons or baryons) in RVMs and comparing it with the same result obtained from latest observations of the CMB anisotropy~\cite{Bond:1993fb,Hu:1994jd,Jungman:1995bz,Hu:2001bc}, new constraints on the dynamics term in RVMs can be obtained, which are independent on previous research on RVMs~\cite{Sola:2007sv,Gomez-Valent:2017idt,Sola:2017jbl,
Rezaei:2019xwo}.

The paper is organized as follows: Sec.~\ref{sec2} is devoted to the review of RVMs. Next, we analyse the energy-momentum tensor conservation and energy conservation of photons in a toy cosmological model and then extend our analysis to a more realistic universe in Sec.~\ref{sec4}. In Sec.~\ref{sec5}, we calculate the particle number density of photons in RVMs, and the similar calculations on the particle number density of baryons are presented in Sec.~\ref{sec6}. Then, based on Secs.~\ref{sec5} and~\ref{sec6}, we study the effect of running vacuum on the baryon-to-photon ratio in Sec.~\ref{sec7}. The last part, Sec.~\ref{sec8}, is a summary and discussion on our results and possible future work.
\section{Running Vacuum Models}
\label{sec2}

The background space-time of the homogeneous and isotropic universe can be described by Friedmann-Robertson-Walker metric:
\begin{eqnarray}\label{metric}
ds^2=-dt^2+a(t)^2 (dx^2+dy^2+dz^2),
\end{eqnarray}
where $a(t)$ is the scale factor, the curvature of the universe has been set as zero, and we employ natural units $c=\hbar=1$.

In the framework of RVMs, the Friedmann equations are similar to the ones in the $\Lambda$CDM model:
\begin{eqnarray}
3H^2=\kappa^2\rho_t,\label{Friedmanneq1}\\
3H^2+2\dot H=-\kappa^2p_t,\label{Friedmanneq2}
\end{eqnarray}
where $H=\dot a/a$ is the Hubble rate, the dot means time derivative, and $\kappa^2=8\pi G$ with $G$ Newton's constant. The parameters $\rho_t$ and $p_t$ represent the total energy density and pressure of all components of the universe, respectively. Since the dominant component is unfixed throughout the evolution of the universe and some of the components can be ignored at particular epochs, the specific forms of $\rho_t$ and $p_t$ rely on the accuracy requirement and the epoch of the universe.

In this work, we consider a class of well described and extensively studied RVMs, in which the vacuum energy density $\rho_\Lambda$ ``runs'' in the form of a function of the Hubble rate based on a renormalization group equation:
\begin{eqnarray}\label{renormalization}
\rho_\Lambda(H)=\frac{\Lambda(H)}{\kappa^2}=\frac{3}{\kappa^2}\left(
c_0+\nu H^2+\alpha\frac{H^4}{H_I^2}\right).
\end{eqnarray}
Note that $c_0$ is a constant with the dimension of energy squared in natural units. Since the current observed value of the Hubble rate is so small, to be consistent with observations, $c_0$ should approximately equal $\frac{1}{3}\Lambda_0$ ($\Lambda_0$ is the value of the cosmological constant in the $\Lambda$CDM model~\cite{Perlmutter:1998np,Aghanim:2018eyx}). The argument $H_I$ can be chosen as the Planck scale or at least below the Planck scale. The dimensionless coefficients $\nu$ and $\alpha$ have been confined to a very narrow range by considering various cosmological phenomena. According to previous research~\cite{Sola:2007sv,Gomez-Valent:2017idt,Sola:2017jbl,
Rezaei:2019xwo}, $\nu$ and $\alpha$ should satisfy $|\nu |<10^{-3}$ and $\alpha\ll1$.

It is believed that $\nu H^2$ has been already much larger than $\alpha\frac{H^4}{H_I^2}$ after BBN~\cite{Lima:2012mu}. In practice, the dynamic characteristics of the vacuum energy is mainly manifested in $\nu H^2$ after BBN, and therefore one can neglect $\alpha\frac{H^4}{H_I^2}$ without conflicting with observations when the problem to be considered has no connection with the early universe. Since our following study does not involve inflation and prior epochs, it is reasonable to discard the subordinate term $\alpha\frac{H^4}{H_I^2}$ in Eq.~(\ref{renormalization}). Hence it is easy to get analytical solutions of Eqs.~(\ref{Friedmanneq1}) and (\ref{Friedmanneq2}) for the radiation-dominated universe by assuming that the main components of the universe are photons and running vacuum. For such simplified cosmological model, Eqs.~(\ref{Friedmanneq1}) and (\ref{Friedmanneq2}) are rewritten as
\begin{eqnarray}
3H^2=\kappa^2(\rho_\Lambda+\rho_r),\label{Friedmanneq11}\\
3H^2+2\dot H=-\kappa^2(p_\Lambda+p_r).\label{Friedmanneq12}
\end{eqnarray}
Since we have already reduced the vacuum energy density (\ref{renormalization}) as $\rho_\Lambda=\frac{3}{\kappa^2}(c_0+\nu H^2)$, one can find directly from Eq.~(\ref{Friedmanneq11}) that the energy density of photons can be written as an explicit function of the Hubble rate. And also if we express the energy density of photons as a function of the scale factor, it is found that
\begin{eqnarray}\label{radia1}
\rho_r(a)=\frac{3}{\kappa^2}(H^2-\nu H^2-c_0)=\rho_{r}(t_0) a^{-4(1-\nu)},
\end{eqnarray}
where we have used the state equation of photons $p_r=1/3\rho_r$ and $p_\Lambda=-\rho_\Lambda$. Note that the integration constant $\rho_r(t_0)$ is not necessarily equal to the current power density of CMB radiation. This solution is only applicable to the radiation-dominated epoch, so the value of $\rho_{r}(t_0)$ should be decided by other cosmological processes in the early universe. Obviously, when $\nu=0$, Eq.~(\ref{radia1}) degenerates into the formula satisfied by photons expanding freely with the universe in the $\Lambda$CDM model.

As for the matter-dominated epoch, one can assume that the universe consists of dark matter, ordinary (baryonic) matter, and running vacuum. So, the Friedmann equations (\ref{Friedmanneq1}) and (\ref{Friedmanneq2}) are given as
\begin{eqnarray}
3H^2=\kappa^2(\rho_\Lambda+\rho_{dm}+\rho_b),\label{Friedmanneq21}\\
3H^2+2\dot H=-\kappa^2(p_\Lambda+p_{dm}+p_b).\label{Friedmanneq22}
\end{eqnarray}
For the sake of simplicity, we set the state equations of baryonic matter and dark matter as $p_b=p_{dm}=0$. In view of the fact that baryonic matter and dark matter have the same state equation and similar energy-momentum tensor (which has only one non-zero component, i.e., energy density), we can regard them as a whole when solving the equations above. From Eqs.~(\ref{Friedmanneq21}) and~(\ref{Friedmanneq22}), it is clear that, the total energy density of baryonic matter and dark matter satisfies
\begin{eqnarray}\label{radia2}
\rho_{dm}+\rho_b=\frac{3}{\kappa^2}[H^2-\nu H^2-c_0]=[\rho_{dm}(t_0)+\rho_{b}(t_0)]a^{-3(1-\nu)}.
\end{eqnarray}
Here, $\rho_{dm}(t_0)+\rho_{b}(t_0)$ is the current total energy density of baryonic matter and dark matter. Apparently, Eq.~(\ref{radia2}) can also degenerate into the standard one in the $\Lambda$CDM model when $\nu=0$. More properties of RVMs can be referred to the literature mentioned above (see Refs.~\cite{Sola:2013gha,Sola:2015rra} for a review).

\section{Energy-momentum tensor conservation and energy conservation}
\label{sec4}

In the $\Lambda$CDM model, photons in the universe travel freely after photon decoupling and the energy-momentum tensor of photons is conserved. Therefore, the total number of photons approximately remains as a constant, which combined with baryon number conservation after BBN leads us to deem that the baryon-to-photon ratio is unchanged after photon decoupling\footnote{Some other views claim that it can even date back to  the BBN epoch because the observations of the light-element abundance indicate that the baryon-to-photon ratio predicated by the primordial nucleosynthesis is close to its current value~\cite{Boesgaard:1985km,Steigman:2007xt,Cyburt:2015mya,
Mathews:2017xht}.}. The conservation of photon number is also based on the requirement that CMB radiation could be regarded as black-body radiation, which results in non-conservation of the internal energy of photons in the universe. In the $\Lambda$CDM model, we can not explain the energy loss of photons due to the expansion of space-time (baryons have no such problem)~\cite{Novello:1989zb,Macleod:2004rd,
Macleod:2005pi}. In this section, we analyse whether the problem can be solved in RVMs.

The internal energy and the number of photons can be expressed as
\begin{eqnarray}
U_r&=&\rho_rV=\frac{\pi^2 k^4}{15c^3\hbar^3}V T^4,\label{internalenergy}\\
N_r&=&\frac{2 k^3\zeta(3)}{\pi^2c^3\hbar^3}V T^3,\label{numberofphotons}
\end{eqnarray}
where $k$ is Boltzmann's constant and $\zeta(n)$ is Riemann zeta function. In the $\Lambda$CDM model, the energy density of CMB satisfies $\rho_r\sim a(t)^{-4}$. Comparing $\rho_r\sim a(t)^{-4}$ with Eq.~(\ref{internalenergy}), it is found that the temperature of CMB is inversely proportional to the scale factor ($T\sim a(t)^{-1}$). Thus photon number density satisfies $n_r=N_r/V\sim T^3\sim a(t)^{-3}$. Because of $V\sim a(t)^3$, $N_r$ is indeed a constant and $U_r\sim T\sim a(t)^{-1}$ decreases with the expansion of the universe. Note that $\rho_r\sim a(t)^{-4}$ can be derived from the conservation of the energy-momentum tensor ($\nabla_\nu T_r^{\nu\mu}=0$). Therefore, in the $\Lambda$CDM model, the energy-momentum tensor conservation of photons ensures the conservation of photon number but leads to a decrease in the internal energy of photons. The energy lost by photons is usually considered to enter the gravitational field or background vacuum energy\footnote{For a single photon propagating in the expanding universe, both its frequency and energy will be reduced due to the expansion of space-time. There is still controversy about the explanation on the energy loss of photons, because the definition of the energy conservation in general relativity is obscure. But, it is for sure that the energy-momentum tensor conservation is not completely equivalent to the energy conservation.}.

Now let us analyse which factors are directly related to the energy loss of photons while the total number of photons remains constant, i.e., $\rho_r\sim a(t)^{-4}$. We consider a specific period of the universe from the scale factor $a(t_1)$ to $a(t_2)$ (the corresponding Hubble rate is $H_1$ and $H_2$, respectively). Supposing that the temperature of CMB is $T_1$ at the moment $a(t_1)$, the number density of photons satisfies
\begin{eqnarray}
n_r(t_1)=\frac{2 k^3\zeta(3)}{\pi^2c^3\hbar^3}T_1^3.\label{numberofphotons2}
\end{eqnarray}
The internal energy and entropy of photons per unit volume (not per comoving volume) are, respectively,
\begin{eqnarray}
\rho_r(t_1)=\frac{\pi^2 k^4}{15c^3\hbar^3} T_1^4,\label{3400}\\
s_r(t_1)=\frac{4\pi^2 k^4}{45c^3\hbar^3} T_1^3.\label{336}
\end{eqnarray}

At the moment $a(t_2)$, the temperature of CMB drops to $T_2$. Since the number of photons remains conserved, the number density can be written as
\begin{eqnarray}\label{36}
n_r(t_2)=\frac{2 k^3\zeta(3)}{\pi^2c^3\hbar^3}T_2^3=\frac{2 k^3\zeta(3)}{\pi^2c^3\hbar^3}\frac{a(t_1)^3}{a(t_2)^3}T_1^3,
\end{eqnarray}
where the second equality is based on $a(t_2)^3T_2^3=a(t_1)^3T_1^3$= Const (the conservation of photon number). The energy lost by CMB per comoving volume is given by
\begin{eqnarray}\label{deltaT}
\Delta u_r=\rho_r(t_2)a(t_2)^3-\rho_r(t_1)a(t_1)^3=\frac{\pi^2 k^4}{15c^3\hbar^3}[T_2^4a(t_2)^3-T_1^4a(t_1)^3].
\end{eqnarray}
Since $a(t_2)^3T_2^3=a(t_1)^3T_1^3$ and $T_2<T_1$, we have $\Delta u_r<0$, which means that as photons expand freely with the universe, the total internal energy of CMB decreases. In the $\Lambda$CDM model, we do not completely figure out where the energy flows. In addition, Eq.~(\ref{deltaT}) manifests that if CMB stays in a state of freely expanding ($\nabla_\nu T_r^{\nu\mu}=0$), the energy loss of CMB within a unit comoving volume is only related to its temperature gradient. Such consequence is comprehensible because the photon gas per unit (comoving) volume is completely characterized by temperature [see Eqs.~(\ref{numberofphotons2}), (\ref{3400}), and (\ref{336})].

Next we study the expansion of photons in RVMs and the change in energy of photons in a unit comoving volume. We focus on the unit comoving volume because it is more practical than the unit volume.  The variation of CMB within the observable universe over a period of time can be reflected in photons per unit comoving volume.

\subsection{Conservation in a toy cosmological model}
\label{sec2.2.1}

We first introduce two special types of conservation in a toy cosmological model, which is only composed of photons and the aforementioned running vacuum. The toy cosmological model is similar to the radiation-dominated universe and it has pedagogical significance for our follow-up research on a realistic universe. The first case is $\nu=0$, and the second corresponds a special value of $\nu$ guaranteeing that the total energy of CMB is a constant. These two special situations represent two types of conservation: energy-momentum tensor conservation and energy conservation.

We first calculate the energy loss ratio of CMB for  $\nu=0$ in the context of the toy cosmological model. The special case of $\nu=0$ indicates that photons can ``freely propagate'' in the universe and the corresponding energy-momentum tensor satisfies $\nabla_\nu T_r^{\nu\mu}=0$. Such property is similar to CMB in the $\Lambda$CDM model. Supposing the temperature of CMB evolves from $T_1$ to $T_2$ with respective scale factors $a(t_1)$ and $a(t_2)$, $\nabla_\nu T_r^{\nu\mu}=0$ results in $a(t_{1})^3T_{1}^3=a(t_2)^3T_2^3$. Therefore, the energy loss ratio of CMB (per unit comoving volume) from $T_1$ to $T_2$ is:
\begin{eqnarray}\label{deltaTrate}
\Gamma_r=\frac{\rho_r(t_1)a(t_1)^3-\rho_r(t_2)a(t_2)^3}
{\rho_r(t_{1})a(t_{1})^3}=\frac{T_{1}-T_2}{T_{1}}.
\end{eqnarray}
It can be found that if the temperature differential of CMB from $a(t_1)$ to $a(t_2)$ is large enough, the energy loss ratio could be close to 1. For example, photons started to ``freely propagate'' after photon decoupling in the $\Lambda$CDM model. It is generally recognized that photon decoupling occurred during recombination about 378,000 years after Big Bang, at a redshift of $z\sim1100$, when the temperature of the universe was $T_1\sim 3000$ K~\cite{Ryden:2003yy,Zyla:2020zbs}. If we take $T_2=2.73$ K, which is the current temperature of CMB~\cite{Mather:1998gm,Fixsen:2009ug,Aghanim:2019ame}, $\Gamma_r$ is approximately equal to 1. Moreover, for a volume equal to the observable universe, the total energy loss of CMB from $T_{1}\sim 3000$ K down to $T_2=2.73$ K is
\begin{eqnarray}\label{39}
\Delta U_r=\Gamma_r \frac{4\pi}{3} R_{1}^3\rho_r(t_1)=\frac{4\pi}{3}
\frac{T_{1}-T_2}{T_{1}}R_{1}^3*\frac{\pi^2 k^4}{15c^3\hbar^3} T_{1}^4\sim 1.6*10^{70}\ \text{J},
\end{eqnarray}
where $R_{1}\sim$ 42 million light-years is the radius of the observable universe at the end of recombination. In this work, we roughly assume that the epoch of photon decoupling is the time at the end of recombination. Note that to get the final consequence, we have recovered the values of all physical quantities above.

Although $\nu=0$ inevitably leads to the energy loss of CMB, the total energy of the toy cosmological model may be conserved by choosing a  proper $c_0$. We also let $a(t_1)$ correspond to the epoch of photon decoupling and $a(t_2)$ be the current scale factor. The energy change of vacuum depends on the density of vacuum and the change in volume of the universe:
\begin{eqnarray}\label{40}
\Delta U_{\Lambda}=\frac{4\pi}{3} \left(R_{2}^3-R_{1}^3\right)\rho_{\Lambda},
\end{eqnarray}
where $R_2$ represents the radius of the current observable universe and $\rho_{\Lambda}=\frac{3}{\kappa^2}c_0$ as $\nu=0$. If $\Delta U_{\Lambda}=\Delta U_r$, $\rho_{\Lambda}$ can be expressed as
\begin{eqnarray}\label{888}
\rho_{\Lambda}=\frac{T_{1}-T_2}{T_{1}}\frac{T_2^3}{T_1^3-T_2^3}
\frac{\pi^2 k^4}{15c^3\hbar^3} T_{1}^4\sim\frac{\pi^2 k^4}{15c^3\hbar^3}T_2^3T_1\sim4.6*10^{-11}\ \text{kg}/(\text{m} \cdot \text{s}^{2}),
\end{eqnarray}
which is less than the current density of vacuum (dark) energy in the $\Lambda$CDM model, $\rho_{\Lambda0}\sim6.3*10^{-10}\ \text{kg}/(\text{m} \cdot \text{s}^{2})$~\cite{Aghanim:2018eyx}. Eq.~(\ref{888}) shows that to keep invariably the energy gained by vacuum equal to the energy lost by photons, vacuum can not be a constant. Regarding this, we will continue to discuss it below.

Next, let us pay attention to another case, i.e., the energy conservation of CMB. First we need to calculate the value of $\nu$ which could maintain the energy conservation of CMB. For the universe composed of photons and vacuum, the Friedmann equations are given by Eqs.~(\ref{Friedmanneq11}) and (\ref{Friedmanneq12}). The energy density of photons satisfies Eq.~(\ref{radia1}). And the energy-momentum tensor conservation of all components leads to
\begin{eqnarray}\label{condition1}
\dot\rho_{\Lambda}+\dot\rho_r+3H(p_{\Lambda}+p_r+
\rho_{\Lambda}+\rho_r)=\dot\rho_{\Lambda}+\dot\rho_r+4H\rho_r=0.
\end{eqnarray}

With Eq.~(\ref{radia1}), we can get the value of $\nu$ directly from the point of view of photons. Since the energy conservation of photons requires
\begin{eqnarray}
\rho_r(t_{2})V(t_{2})=\rho_r(t_{1})V(t_{1}),
\end{eqnarray}
it is found that the parameter $\nu$ should be equal to $\frac{1}{4}$, which can be also derived from $a(t_{2})^3T_{2}^4=a(t_1)^3T_1^4$.

In order to verify whether the value of $\nu$ also guarantees the energy conservation of vacuum, we need to solve Eq.~(\ref{Friedmanneq3}), obtain the Hubble rate $H$ as a function of $a(t)$, and then take it back to $\rho_{\Lambda}(H)$ to get $\rho_{\Lambda}(a(t))$. Finally, if $\rho_{\Lambda}(t_{2})V(t_{2})=\rho_{\Lambda}(t_{1})V(t_{1})$ is true for $\nu=\frac{1}{4}$, it proves that $\nu=\frac{1}{4}$ can indeed ensure the energy conservation of photons and vacuum, respectively, while their total energy-momentum tensor satisfies the conservation condition (\ref{condition1}). Combining Eq.~(\ref{condition1}) and $\rho_r(a)=\rho_{r}(t_0) a^{-4(1-\nu)}$, we have
\begin{eqnarray}\label{condition0}
\frac{d\rho_{\Lambda}}{da}+\frac{d\rho_{r}}{da}+
4\frac{\rho_{r}}{a}=\frac{d\rho_{\Lambda}}{da}+4\nu \rho_{r}(t_0)a^{-5+4\nu}=0.
\end{eqnarray}
The solution can be expressed as $\rho_{\Lambda}(a)=\frac{\nu}{1-\nu}\rho_r(t_0)a^{-4(1-\nu)}+
\frac{3c_0}{1-\nu}=\frac{\nu}{1-\nu}\rho_r+\frac{3c_0}{1-\nu}$. Taking $\rho_{\Lambda}(a)$ into $\rho_{\Lambda}(t_{2})V(t_{2})=\rho_{\Lambda}(t_{1})V(t_{1})$ yields
\begin{eqnarray}\label{315}
\frac{\nu}{1-\nu}\rho_r(t_{2})V(t_{2})+\frac{3c_0}{1-\nu}V(t_{2})=
\frac{\nu}{1-\nu}\rho_r(t_{1})V(t_{1})+\frac{3c_0}{1-\nu}V(t_{1}).
\end{eqnarray}
Since $\rho_r(t_{2})V(t_{2})=\rho_r(t_{1})V(t_{1})$ and $V(t_{2})\neq V(t_{1})$, Eq.~(\ref{315}) is valid only when the constant term $c_0$ of the vacuum energy is zero. Thereupon, the evolution of $\rho_{\Lambda}=\frac{\nu}{1-\nu}\rho_r(t_0)a^{-4(1-\nu)}$ along with the scale factor is exactly the same as the situation of photons [see Eq.~(\ref{radia1})].

Since the temperature of CMB satisfies $a(t_{2})^3T_{2}^4=a(t_1)^3T_1^4$, the entropy change of CMB from the epoch of photon decoupling to the current day is given as
\begin{eqnarray}
\Delta S=\frac{4\pi^2 k^4}{45c^3\hbar^3}[V(t_{2})T_{2}^3 -V(t_{1})T_{1}^3]>0,
\end{eqnarray}
which conforms with the second law of thermodynamics. The entropy of the current CMB has increased by several orders of magnitude compared to its value at the end of recombination on account of $T_{1}\gg T_{2}$. As far, we have proved that both of the energy conservation of photons and vacuum can be guaranteed in the toy cosmological model as long as $\nu=\frac{1}{4}$ and $c_0$ is vanishing.

Finally, we summarize briefly the characteristics of these two special kinds of conservation in above toy cosmological model. When CMB freely expands ($\nu=0$ and so $\nabla_{\mu\nu}T_r^{\mu}=0$), the corresponding photon number and entropy keep fixed. In this case, not only is the energy of CMB non-conserved, but also the total energy of the universe can not be conserved. So, in this toy cosmological model the energy-momentum tensor conservation of CMB inevitably causes the non-conservation of the total energy of the universe. For another special case that CMB maintains energy conservation, based on $\rho_r(t_{2})V(t_{2})=\rho_r(t_{1})V(t_{1})$, it can be found that the value of the parameter $\nu$ must be $\frac{1}{4}$. The energy density of CMB evolves with the scale factor in the same way as dust. And if the constant term $c_0$ of the vacuum energy equals zero, the energy of vacuum is also conserved. Therefore, $\rho_{\Lambda}=\frac{3}{4\kappa^2} H^2$ can guarantee both of the energy conservation of photons and vacuum. As for the rest of $\nu$ and $c_0$, neither the energy-momentum tensor conservation of photons nor the energy conservation of photons can be guaranteed. And it can be seen from Eq.~(\ref{315}) that there is no other $\nu$ that keeps the total energy of the universe conserved. Next, let us study a more realistic cosmological model and discuss whether RVMs could balance observational data and the energy conservation of the universe.

\subsection{Conservation in a realistic cosmological model}
\label{sec2.2.3}
In this section, we study the energy-momentum tensor conservation and energy conservation of CMB in a realistic universe with running vacuum. We suppose that vacuum is only coupled to photons and the other components of the universe satisfy their own energy-momentum tensor conservation and also energy conservation.

In the realistic cosmological model, if there is an interaction\footnote{The vacuum being a cosmological constant does not mean that vacuum can not interact with photons or other components of the universe. With the expansion of space-time, the total energy of vacuum within the observable universe is increasing although the energy density of vacuum is a constant. Therefore, it is entirely reasonable to believe that the increased energy of vacuum is caused by the interaction between vacuum and other substances.} between photons and vacuum, it should start when photons emerged in the universe (the temperature of the universe was about $T_e\sim10^{16}$ K at the end of the electroweak
epoch). There should also be a cutoff temperature $T_c$, at which photons and vacuum are decoupled. At present, we can not evaluate $T_c$ from the first principle or existing knowledge, so we temporarily deem that $T_c$ just need satisfy $0<T_c<T_e$. Here, we suppose that the cutoff temperature satisfies $0<T_c<T_0=2.73$ K, which indicates that there is still an interaction between photons and vacuum.

We have already mentioned that photons ``freely propagate'' after the recombination epoch. Therefore, when studying the properties of the interaction between photons and running vacuum, we only consider the period of the universe after recombination, which helps to exclude the interaction between photons and other substances. From the end of recombination to the current day, the realistic universe is no longer dominated by radiation, but dust (or baryonic matter), dark matter, and vacuum (or dark energy). All components of the universe are not negligible except for neutrinos, so the proper Friedmann equations should be given as
\begin{eqnarray}\label{Friedmanneq3}
3H^2=\kappa^2(\rho_\Lambda+\rho_{dm}+\rho_b+\rho_{r}),\label{Friedmanneq31}\\
3H^2+2\dot H=-\kappa^2(p_\Lambda+p_{dm}+p_b+p_r).\label{Friedmanneq32}
\end{eqnarray}
Hereafter, we just set $\kappa^2=1$ for simplicity. For dust and dark matter with a vanishing pressure, energy conservation is equivalent to energy-momentum tensor conservation ($\nabla_{\mu\nu}T^{\nu}_{(b,dm)}=0\longrightarrow \dot \rho_{(b,dm)}+3H \rho_{(b,dm)}=0\longrightarrow V\rho_{(b,dm)}\equiv\text{Const}$). Therefore, the total energy-momentum tensor conservation of photons and vacuum is still Eq.~(\ref{condition1}), which yields the energy density of photons as
\begin{eqnarray}\label{314}
\rho_r(a)&=&\rho_{r0} a^{-4(1-\nu)}+\frac{3\nu [\rho_b(t_0)+\rho_{dm}(t_0)] }{1-4\nu} a^{-3}\ \ \ \ \ \ \ \ (\nu\neq\frac{1}{4})\nonumber\\
\rho_r(a)&=&\rho_r({t_0}) a^{-3}+\frac{3[\rho_b(t_0)+\rho_{dm}(t_0)]}{4}a^{-3}\log a\ \ \ \ \ \ (\nu=\frac{1}{4}).
\end{eqnarray}
Note that $\rho_b(t_0)$ and $\rho_{dm}(t_0)$ are the current values of baryon density and dark matter density, respectively. $\rho_r({t_0})$ is the current energy density of CMB, but $\rho_{r0}\neq\rho_r(t_0)$, which is decided by $\rho_r({t_0})$:
 \begin{eqnarray}\label{20}
\rho_{r0}=\rho_r(t_0)-\frac{3\nu [\rho_b(t_0)+\rho_{dm}(t_0)] }{1-4\nu}.
\end{eqnarray}

Combining Eqs.~(\ref{condition1}) and (\ref{314}), we have
\begin{eqnarray}\label{condition2}
\frac{d\rho_{\Lambda}}{da}+\frac{d\rho_{r}}{da}+
4\frac{\rho_{r}}{a}=\frac{d\rho_{\Lambda}}{da}+4\nu \rho_{r0}a^{-4(1-\nu)-1}+\frac{3\nu[\rho_b(t_0)+\rho_{dm}(t_0)]}
{1-4\nu}a^{-4}=0.
\end{eqnarray}
The solution of the vacuum density can be expressed as
\begin{eqnarray}\label{317}
\rho_{\Lambda}(a)&=&\frac{\nu}{1-\nu}\rho_{r0}a^{-4(1-\nu)}+
\frac{\nu[\rho_b(t_0)+\rho_{dm}(t_0)]}{1-4\nu}a^{-3}+
\frac{3}{1-\nu}c_0\ \ (\nu\neq\frac{1}{4}\ \  \text{and}\ \ \nu\neq1)\nonumber\\
\rho_{\Lambda}(a)&=&H^2_0a^{-3}+
\frac{1}{4}[\rho_b(t_0)+\rho_{dm}(t_0)]a^{-3}\log a+ 4c_0\ \ (\nu=\frac{1}{4})\nonumber\\
\rho_{\Lambda}(a)&=&-\frac{\rho_b(t_0)+\rho_{dm}(t_0)}{3}a^{-3}
-4\rho_{r0} \log a \ \ (\nu=1),
\end{eqnarray}
where $H^2_0=\frac{1}{3}[\rho_b(t_0)+\rho_{dm}(t_0)
+\rho_r(t_0)]$. For the solution with $\nu=1$, we have $\rho_{r0}=-3 c_0$.

With the solutions, we can study the two special types of conservation of photons in the realistic universe. When $\nu=0$ (the energy-momentum tensor conservation of photons), RVMs revert to the $\Lambda$CDM model and photons could ``freely expand'' after photon decoupling. The scale factor $a(t_{re})$ of the universe at the end of recombination can be estimated by the fact that freely propagating photons satisfy $a(t)^3T^3=$ Const. Since $T_{re}$ is about $3000$ K and $a(t_{re})^3T_{re}^3=a(t_0)^3T_0^3$, according to the current radius of the observable universe ($R_{0}\sim$ 46 billion light-years), the radius of the universe corresponding to the observable universe at the end of recombination is $R_{re}\sim$ 42 million light-years. The radius is independent on the components of the universe and so is consistent with the result (see $R_1$ in the previous section) obtained in the toy cosmological model. Although the estimation on $R_{re}$ is rough, it is still pretty much the same as the result obtained in the $\Lambda$CDM model, and such precision has no impact on our subsequent discussions. Setting the scale factor of the current universe as $a(t_0)=1$ yields $a(t_{re})\sim 9.1*10^{-4}$.

Note that when $\nu=0$, the solution of $\rho_r$ is consistent with the one in the $\Lambda$CDM model and also the one in the toy cosmological model, so we can use directly the previous calculations and results. It is shown that for the process that the temperature of CMB drops from 3000 K to 2.73 K, the energy loss of the observable universe is still Eq.~(\ref{39}). In the same process, taking into account the current vacuum energy density, $\rho_{\Lambda0}\sim6.3*10^{-10}\ \text{kg}/(\text{m} \cdot \text{s}^{2})$~\cite{Aghanim:2018eyx}, the energy change of vacuum is
\begin{eqnarray}\label{399}
\Delta U_{\Lambda}=\frac{4\pi}{3} \left(R_{0}^3-R_{re}^3\right)\rho_{\Lambda0}\sim2.2*10^{71}\ \text{J}.
\end{eqnarray}
Comparing Eqs.~(\ref{39}) and~(\ref{399}), one can find that these two values are not accordant and the gap is more than tenfold, that is to say, the energy lost by photons is only about a tenth of the energy obtained by vacuum. The previous toy cosmological model indicates that when photons expand freely, the total energy of the universe can not be maintained. It now appears that the conclusion also applies to the $\Lambda$CDM model. Moreover, since the energy-momentum tensor conservation of photons is also equivalent to photon number conservation, the entropy of CMB is still unchanging, which is about $10^{89}\ k$~\cite{Kolb:1990vq,Egan:2009yy}. In the $\Lambda$CDM model, we can not explain why the entropy of CMB is so large~\cite{Kolb:1990vq}.

In the toy cosmological model, we found that both the photon energy and the vacuum energy can be conserved in RVMs with $\nu =1/4$ and $c_0 = 0$, so the total energy of the universe is conserved. In the realistic universe, let us check if there are such $\nu$ and $c_0$. The energy conservation of CMB signifies $a(t_{2})^3T_{2}^4=a(t_1)^3T_1^4$, with which the radius of the observable universe at the end of recombination can be obtained as $R_{re}\sim 4.1$ million light-years. And then the corresponding scale factor is $a(t_{re})\sim8.8*10^{-5}$. Note that the influence of running vacuum on the scale factor is still within the acceptable range~\cite{Felten:1986zz,Overduin:1998zv}.

Following the steps above, one may get the value of $\nu$ directly from Eq.~(\ref{314}). Taking Eqs.~(\ref{314}) and (\ref{317}) into $\rho_r(t_{re})V(t_{re})=\rho_r(t_{0})V(t_{0})$, we have
\begin{eqnarray}\label{25}
a(t_{re})^{-1+4\nu}&=&a(t_0)^{-1+4\nu}\ \ \ \ (\nu\neq\frac{1}{4})\nonumber\\
\log[a(t_{re})]&=&\log[a(t_{0})]\ \ \ \ \ \ (\nu=\frac{1}{4}).
\end{eqnarray}
Unfortunately, for the solution with $\nu\neq\frac{1}{4}$ [see Eq.~(\ref{314})], the condition guaranteeing the energy conservation of photons is exactly $\nu=\frac{1}{4}$, while the solution with $\nu=\frac{1}{4}$ requires the scale factor to be a constant. Therefore, there is no $\nu$ that could render Eq.~(\ref{25}) true due to the existence of other substances, i.e., $\rho_b(t_0)+\rho_{dm}(t_0)\neq0$.

Although there is no $\nu$ keeping photon energy conserved, it is likely that there exists a special $\nu$ sustaining the total energy conservation of the universe, which means $[\rho_r(t_{re})+\rho_{\Lambda}(t_{re})]V(t_{re})=
[\rho_r(t_{0})+\rho_{\Lambda}(t_{0})]V(t_{0})$. It is found that the energy conservation of the universe is valid only when $\nu=1$ and $\rho_{r0}=-3c_0=0$. Reviewing Eq.~(\ref{314}), the energy density of photons is negative in case of $\nu=1$ and $\rho_{r0}=-3c_0=0$, which is obviously unacceptable.

By now, we have discussed two special couplings between running vacuum and CMB in the context of a realistic cosmological model. Our research shows that it is impossible for CMB to satisfy energy conservation in RVMs due to $\rho_b(t_0)+\rho_{dm}(t_0)\neq0$. Even replacing the independent energy conservation of CMB with the energy conservation of the whole universe, it can not be achieved by choosing some special values of $\nu$ and $c_0$. Next, we discuss the impact of running vacuum on the baryon-to-photon ratio. We first study the properties of photons and baryons in RVMs, respectively.

\section{Photons in RVMs}
\label{sec5}

CMB is a powerful tool and indicator for studying the evolution of the universe. In different cosmological models, the properties of CMB may be completely different. Combining observations and reasonable inferences, one can put forward effective constraints on cosmological models and even eliminate fallacious cosmological models. In RVMs, running vacuum can influence the evolution of certain components of the universe by a non-minimal coupling. If running vacuum could be coupled to CMB, after photons are decoupled from other matter, CMB will be mainly controlled by the expansion of the universe (scale factor) and running vacuum. According to the current observations of CMB and the temperature estimation of CMB at the end of recombination from the CMB anisotropy, the effective parameter space of running vacuum may be limited to a narrow range. In this section, we study the properties of CMB after photon decoupling in RVMs. Considering the value of $\nu$ in RVMs has been limited to $|\nu |<10^{-3}$ in previous researches~\cite{Sola:2007sv,Gomez-Valent:2017idt,Sola:2017jbl,
Rezaei:2019xwo}, our following research is only relevant to the solution for $\nu<\frac{1}{4}$ in Eqs.~(\ref{314}) and (\ref{317}).

In order to make our discussion more clear and reasonable, let us sort out the known facts about CMB and the observable universe in advance. First of all, the current temperature of CMB ($T_0=2.73$ K) and the radius of the observable universe ($R_{0}\sim$ 46 billion light-years) are given by precise observations, which can not be violated in any model of the universe. Secondly, according to the theory of scattering describing the collision between photons and atoms (electrons), the decoupling temperature of CMB is determined around 3000 K~\cite{Ryden:2003yy,Zyla:2020zbs}. Finally, the size of the observable universe at the epoch of photon decoupling is indistinct (i.e., the value of the scale factor), which affects our evaluation on the age\footnote{Generally, it is considered as 378,000 years in the $\Lambda$CDM model.} and the baryon-to-photon ratio of the universe at the epoch of photon decoupling. Based on these facts, we can calculate and analyze reasonably the impact of running vacuum on CMB in RVMs.
The number of photons per unit comoving volume (not unit volume) is the most concerned issue in this section.

Taking $\rho_{r0}$ [see Eq.~(\ref{20})] into Eq.~(\ref{314}), since the energy of photons per unit volume satisfies Eq.~(\ref{3400}), the relation between the temperature of CMB and the scale factor is given as
\begin{eqnarray}\label{77788}
\left(\rho_r(t_0)-\frac{3\nu [\rho_b(t_0)+\rho_{dm}(t_0)] }{1-4\nu}\right) a^{-4(1-\nu)}+\frac{3\nu [\rho_b(t_0)+\rho_{dm}(t_0)] }{1-4\nu} a^{-3}=\frac{\pi^2 k^4}{15c^3\hbar^3} T^4.
\end{eqnarray}
When $T=T_0=2.73$ K and $a(t_0)=1$, $\rho_r(t_0)\sim4.0*10^{-14}\text{J/m}^3$ is not related to $\nu$. When $\nu=0$ (the $\Lambda$CDM model), $a(t_{re})$ is approximately given as $9.1*10^{-4}$ with redshift $z\sim1100$.

\begin{figure*}
\begin{center}
\subfigure[]{\label{pic1}
\includegraphics[width=7cm,height=4.5cm]{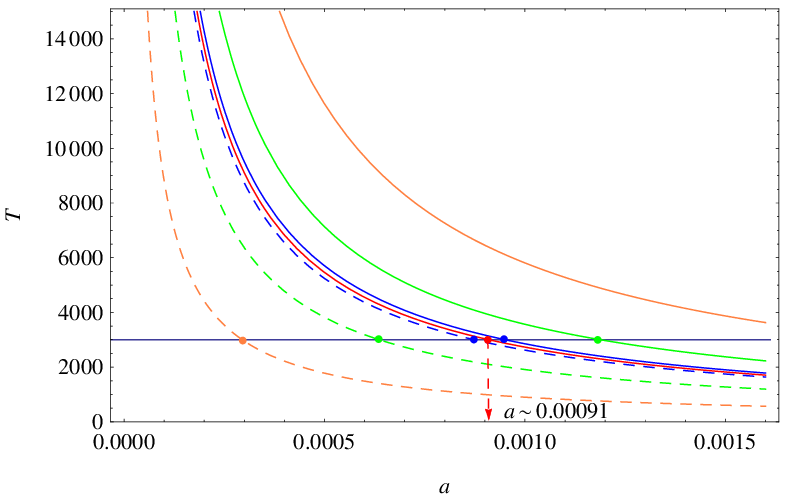}}
\subfigure[]{\label{pic15}
\includegraphics[width=7cm,height=4.5cm]{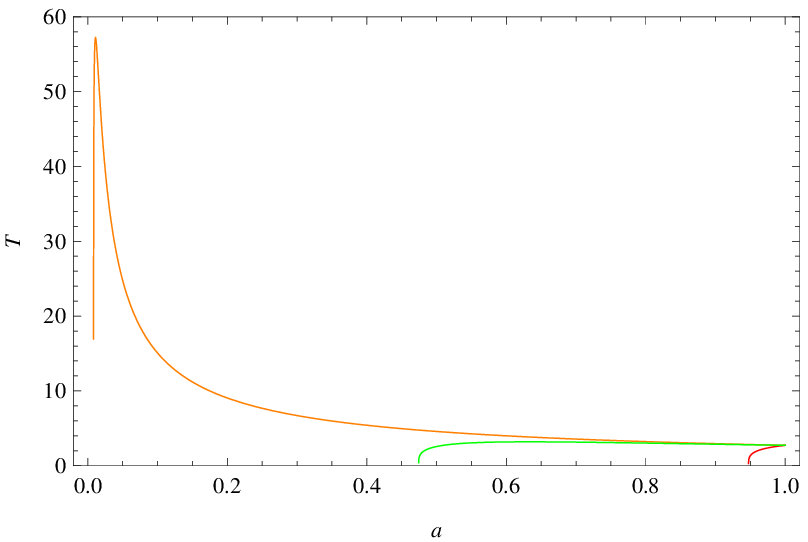}}
\subfigure[]{\label{pic2}
\includegraphics[width=7cm,height=4.65cm]{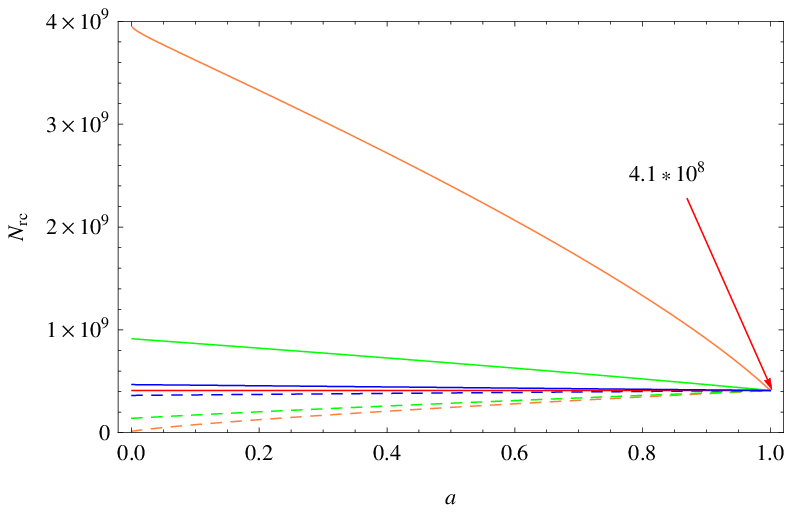}}
\subfigure[]{\label{picex}
\includegraphics[width=7cm,height=4.5cm]{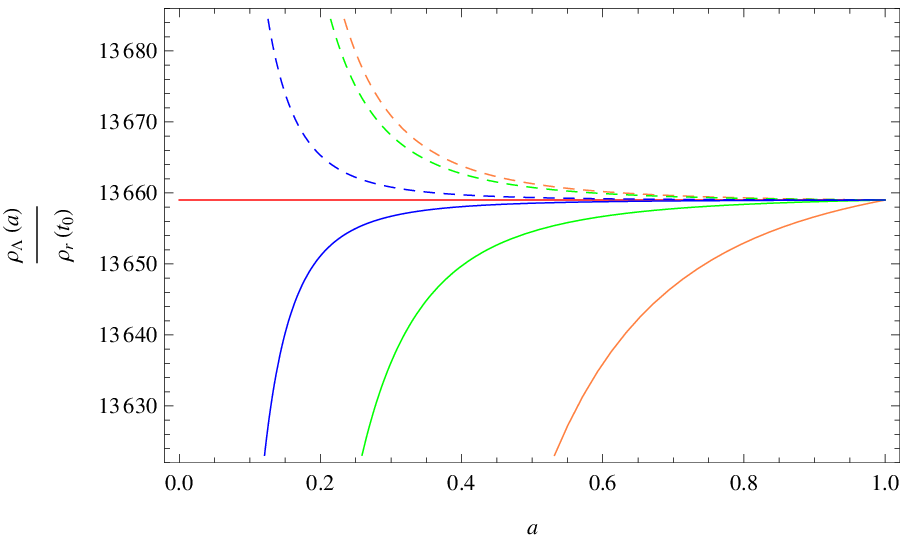}}
\end{center}
\caption{Both of the upper panels are the temperature of CMB evolving with the scale factor. The under left panel is the number of photons per unit comoving volume. The under right panel is the energy density of running vacuum, which has been plotted in a dimensionless manner. For Figs.~\ref{pic1}, \ref{pic2}, and \ref{picex}, seven cases of the parameter $\nu$ are presented: $\nu=-10^{-3}$ (orange-solid), $\nu=-10^{-4}$ (green-solid), $\nu=-10^{-5}$ (blue-solid), $\nu=0$ (red-solid), $\nu=8*10^{-6}$ (blue-dashed), $\nu=4*10^{-5}$ (green-dashed), and $\nu=5.2*10^{-5}$ (orange-dashed). In Fig.~\ref{pic15}, three cases of the parameter $\nu$ are presented: $\nu=10^{-3}$ (red), $\nu=10^{-4}$ (green), and $\nu=5.3*10^{-5}$ (orange). For all plots, the proportions of each component in the current universe are: $26.8\%$ (DM), $4.9\%$ (Baryon), $0.005\%$ (CMB), and $68.295\%$ (Vacuum).}
\end{figure*}

With Eq.~(\ref{77788}), it is found that the sign of $\nu$ has a decisive influence on the value of the scale factor [see Fig.~\ref{pic1}]. All temperature lines in Fig.~\ref{pic1} converge to the point ($a=1$, $T=2.73$ K), and the intersections of the temperature lines with the horizontal line $T=3000$ K are the values of the scale factor at the end of recombination for different $\nu$ (see Table~\ref{table1} for detail). When $\nu>0$, the scale factor at the end of recombination is less than the estimated value in the $\Lambda$CDM model, and the scale factors resulted from $\nu<0$ have exactly reverse effects. Overall, the scale factor of recombination increases with $\nu$.

Our research shows that within the reference range of $|\nu |<10^{-3}$, the evolution of CMB can indeed provide extra strict constraints on the range of $\nu$ if vacuum is coupled to photons. When $\nu>0$, we find that the upper limit of $\nu$ is about $5.25*10^{-5}$, and the larger $\nu$ would result in non-physical results of CMB [see Fig.~\ref{pic15}]. But when $-10^{-3}<\nu<0$, the energy density of photons is always normal.

The number of photons per unit comoving volume is given as
\begin{eqnarray}\label{4200}
N_{rc}=\rho_r(a)\ a^3=
r_A\left(\frac{T_0^4[a^{4\nu}(1-19024\nu)+19020\nu a]}
{a^4(1-4\nu)}\right)^{3/4}\ a^3,
\end{eqnarray}
where $r_A=\frac{2k^3\zeta(3)}{\pi^2c^3\hbar^3}\sim 2.02*10^7$~(K$\cdot$ m)$^{-3}$. For some special values of the parameter $\nu$, the evolution of $N_{rc}$ is shown in Fig.~\ref{pic2}. Only when $\nu=0$, $N_{rc}$ does not evolve with the temperature change of CMB [the red-solid line in Fig.~\ref{pic2}], which is exactly what happens in the $\Lambda$CDM model. Since the parameter $\nu$ does not change the temperature of CMB at present and also the energy density of photons, when $a=1$, all cases correspond to the current number density of photons in the universe: $4.1*10^{8}\ \text{m}^{-3}$ [see Fig.~\ref{pic2}]. We also noticed that although the number density of photons at recombination is the same (it is a function of temperature and independent of $\nu$ and the scale factor), the number of photons per unit comoving volume can differ more than two orders of magnitude due to the parameter $\nu$. Even for the case of $\nu\sim-10^{-5}$, the gap, comparing with the $\Lambda$CDM model, could be close to $15\%$ (see more detail in Table~\ref{table1}).

Finally, the energy density of running vacuum is shown in Fig.~\ref{picex}, which evolves monotonously with the scale factor (monotone decreasing for $\nu>0$ and monotone increasing for $\nu<0$).

\begin{table}[htbp]
	\centering  
	\caption{Scale factor and photon number per unit comoving volume at the end of recombination}  
	\label{table1}  
	\begin{tabular}{|c|c|c|c|c|c|c|c|c}
		\hline  
		& & & & & & &\\[-6pt]  
		$\nu$ & $-10^{-3}$ & $-10^{-4}$ & $-10^{-5}$ & 0 & $8*10^{-6}$ & $4*10^{-5}$ & $5.2*10^{-5}$ \\  
		\hline
		& & & & & & &\\[-6pt]  
		$a(t_{re})$ & 0.00193 & 0.00118 & 0.00095 & 0.00091 & 0.00087 & 0.00064 & 0.00029\\
		\hline
		& & & & & & &\\[-6pt]  
		$N_{rc}({t_{re}})$ & $3.9*10^9$ & $9.1*10^8$ & $4.7*10^8$ & $4.1*10^8$ & $3.6*10^8$ & $1.4*10^8$ & $1.4*10^7$\\
		\hline
	\end{tabular}
\end{table}

\section{Baryons in RVMs}
\label{sec6}

Since the number of baryons in the universe is a billion times less than the number of photons, the variation of the baryon-to-photon ratio is more sensitive to the variation of the number density of baryons. In this section, we discuss the possible impact of running vacuum on baryons in the universe. Similar to the previous approach, when baryons are coupled to vacuum, we assume that the energy-momentum tensor of CMB maintains independently conservation.

For the coupling between running vacuum and baryons, the corresponding Friedmann equations are still Eqs.~(\ref{Friedmanneq3}) and (\ref{Friedmanneq32}). Since the solutions for photons and running vacuum with respect to the scale factor are the same as the ones in the $\Lambda$CDM model, we can immediately get the energy density of baryons being satisfied with
\begin{eqnarray}\label{condition8}
\frac{d\rho_{\Lambda}}{da}+\frac{d\rho_{b}}{da}+
3\frac{\rho_{b}}{a}=-\frac{3\nu}{a}\left[\frac{4}{3}\rho_r(t_0)a^{-4}
+\rho_{dm}(t_0)a^{-3}+\rho_b\right]+\frac{d \rho_b}{d a}+3\frac{\rho_b}{a}=0.
\end{eqnarray}
The solution is given as
\begin{eqnarray}\label{9999}
\rho_b(a)&=&\rho_{b0}a^{-3(1-\nu)}-\rho_{dm}(t_0)a^{-3}
-\frac{4\nu\rho_r(t_0)}{1+3\nu}a^{-4}\ \ \ \ (\nu\neq-\frac{1}{3})\nonumber\\
\rho_b(a)&=&\rho_{b1}a^{-4}-\rho_{dm}(t_0)a^{-3}
-\frac{4\rho_r(t_0)}{3}a^{-4}\log a\ \ \ \ \ \ (\nu=-\frac{1}{3}).
\end{eqnarray}
Noth that $\rho_{b0}=\rho_b(t_{0})+\rho_{dm}(t_0)+
\frac{4\nu\rho_r(t_0)}{1+3\nu}$ and $\rho_{b1}=\rho_b(t_{0})+\rho_{dm}(t_0)$, which insure that when $\nu=0$, $\rho_b(a)$ could revert to the solution in the $\Lambda$CDM model. And the solution of running vacuum is
\begin{eqnarray}\label{31722}
\rho_{\Lambda}(a)&=&\frac{\nu\rho_r(t_0)}{1+3\nu}a^{-4}+
\frac{\nu \rho_{b0}}{1-\nu}a^{-3(1-\nu)}+\frac{3}{1-\nu}c_0\ \  (\nu\neq-\frac{1}{3}\ \ \text{and}\ \ \nu\neq1)\nonumber\\
\rho_{\Lambda}(a)&=&-\frac{1}{4}[\rho_r(t_0)+\rho_{b1}]a^{-4}+
\frac{1}{3}\rho_r(t_0)a^{-4}\log a +\frac{9}{4}c_0\ \ (\nu=-\frac{1}{3})\nonumber\\
\rho_{\Lambda}(a)&=&4\rho_r(t_0)a^{-4}-3\rho_{b0}\log a\ \ (\nu=1),
\end{eqnarray}
where $\rho_{b0}=-3c_0$ and the corresponding integration constant has been set to 0. With these solutions, we can study the evolution of baryons after recombination. Also, since Joan Sol$\grave{\text{a}}$ et al. have pointed out, with multiple observations in cosmology, $|\nu |<10^{-3}$~\cite{Sola:2007sv,Gomez-Valent:2017idt,Sola:2017jbl,
Rezaei:2019xwo}, we only discuss the impact caused by running vacuum on the number density of baryons within the scope of $|\nu |<10^{-3}$.

\begin{figure*}
\begin{center}
\subfigure[]{\label{pic3}
\includegraphics[width=7cm,height=4.5cm]{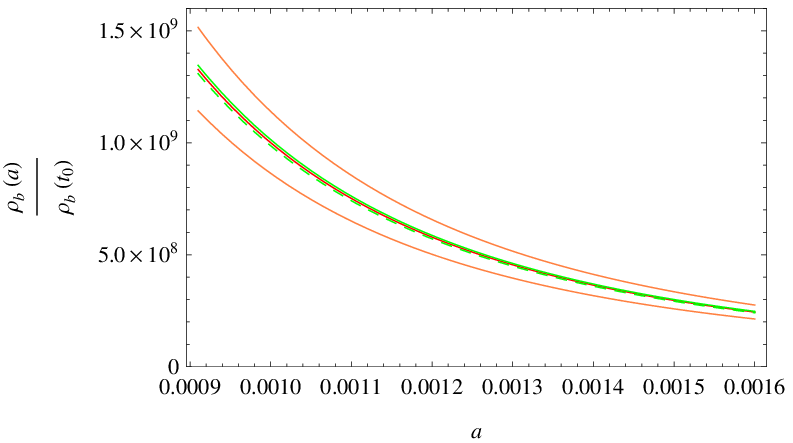}}
\subfigure[]{\label{pic4}
\includegraphics[width=7cm,height=4.5cm]{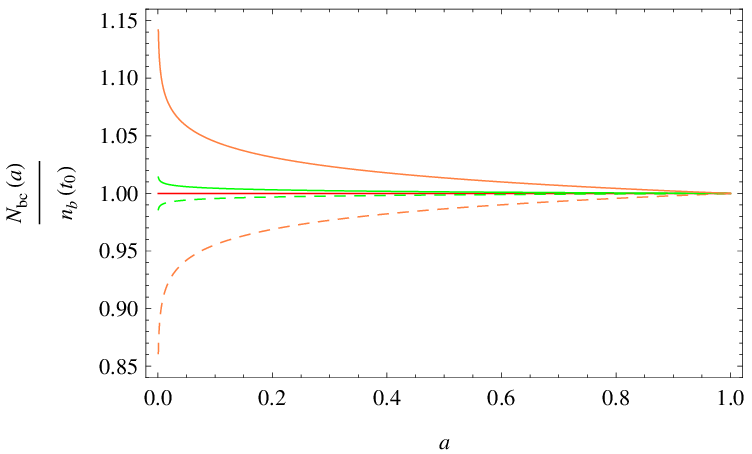}}
\subfigure[]{\label{pic5}
\includegraphics[width=7cm,height=4.5cm]{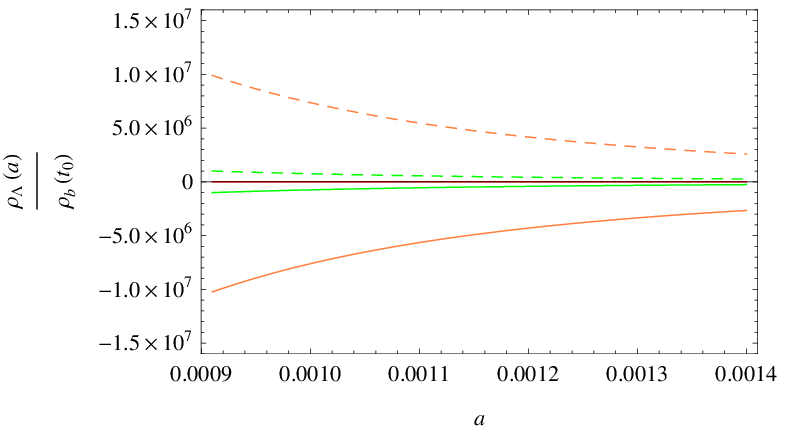}}
\end{center}
\caption{The upper left panel is the energy density of baryons. The upper right panel is the number of baryons per unit comoving volume. The under panel is the energy density of running vacuum. All plots have been plotted in a dimensionless manner. Five cases of the parameter $\nu$ are presented in all plots: $\nu=-10^{-3}$ (orange-solid), $\nu=10^{-4}$ (green-solid), $\nu=0$ (red-solid), $\nu=10^{-3}$ (orange-dashed), and $\nu=10^{-4}$ (green-dashed). The current number of baryons per unit comoving volume is about $0.2$. The proportions of each component in the current universe are: $26.8\%$ (DM), $4.9\%$ (Baryon), $0.005\%$ (CMB), and $68.295\%$ (Vacuum).}
\end{figure*}

Suppose there exists a baryon-generation (annihilation) process in the universe and the energy source (sink) is running vacuum\footnote{It seems that such a process is most likely to exist in the early universe with high energy and strong gravitational field. The known possible baryon number non-conservation processes are basically ruled out in the low-energy late universe. However we do not rule out the possibility that the interaction between running vacuum and baryons, which could be extremely weak and hardly be detected. Later we will simply estimate the incidence of the interaction in RVMs.}. According to the above analysis on the freely expanding photons, it can be known that the approximate value of the scale factor is 0.00091 at the end of recombination. The baryon density of the current universe is approximately $4.2*10^{-28}\ \text{kg}/\text m^3$. Therefore, the baryons density at the end of recombination can be obtained from Eq.~(\ref{9999}), which is plotted in Fig.~\ref{pic3} with five sets of $\nu$'s. When $|\nu |<10^{-5}$, the baryon density is almost hard to deviate from the result ($\nu=0$) in the $\Lambda$CDM model. Even if $|\nu |$ takes the maximum and minimum values of the feasible range $|\nu |<10^{-3}$, the deviation of the baryon density from the result in the $\Lambda$CDM model is still negligible. At the end of recombination, the maximum deviation is about $15\%$ (see Table~\ref{table2}). As $\nu<0$, running vacuum presents obviously a state of negative energy, which actually should also appear in Fig.~\ref{picex} when $\nu<0$ and $a(t)\ll1$. Unlike baryons, running vacuum is sensitive to the value of $\nu$ at the end of recombination. Even a tiny deviation of $\nu$ from $\nu=0$ can cause an order of magnitude difference in the vacuum energy density [see Fig.~\ref{pic5}]. When $\nu=0$, $\rho_\Lambda=3c_0\sim6.3*10^{-10}\ \text{kg}/(\text{m} \cdot \text{s}^{2})$ is the current vacuum energy density [see the red-solid line in Fig.~\ref{pic5}].

\begin{table}[htbp]
	\centering  
	\caption{Baryon density and baryon number per unit comoving volume at the end of recombination}
	\label{table2}  
	\begin{tabular}{|c|c|c|c|c|c|c|c|c|c|c|c}
		\hline  
		& & & & & & &\\[-6pt]  
		$\nu$ & $-10^{-3}$ & $-10^{-4}$ & $-10^{-5}$ & 0 & $10^{-5}$ & $10^{-4}$ & $10^{-3}$\\  
		\hline
		& & & & & & &\\[-6pt]  
		$10^{-9}\rho_{b}(t_{re})/\rho_{b}(t_{0})$ & $1.143$ & $1.308$ & $1.325$ & $1.327$ & $1.329$ & $1.346$ & $1.515$\\
		\hline
& & & & & & &\\[-6pt]  
		$n_b(t_0)^{-1}N_{bc}(t_{re})$ & $0.86113$ & $0.98599$ & $0.99860$ & $1$ & $1.00140$ & $1.01404$ & $1.14175$ \\
		\hline
	\end{tabular}
\end{table}

Now we analyse the number density of baryons, which can be expressed as
\begin{eqnarray}\label{00000}
\rho_b(a)=n_b(a) m_0 c^2,
\end{eqnarray}
where $m_0$ is the mass of a single baryon, $n_b(a)$ is the number density of baryons, and $c$ is the speed of light. According to the current observations, $n_b(t_0)\sim2.5*10^{-7}$ atoms$\cdot$cm$^{-3}$ is a relatively accurate value\footnote{ The estimated value of $n_b(t_0)$ from BBN is in the range of $(2.4*10^{-7}\sim2.7*10^{-7})$ atoms$\cdot$cm$^{-3}$ ($95\%$ CL)~\cite{Tanabashi:2018oca}.}. If $m_0$ is a constant, $n_b(t_{re})=\frac{\rho_b(t_{re})}{\rho_b(t_0)}n_b(t_0)$. The number of baryons per unit comoving volume satisfies
\begin{eqnarray}\label{0033000}
N_{bc}(a)=n_b(a) a^3=\frac{\rho_b(a)}{\rho_b(t_0)}n_b(t_0)a^3,
\end{eqnarray}
where $\rho_b(a)$ refers to  Eq.~(\ref{9999}). The evolution of $N_{bc}(a)$ with the scale factor is given by Fig.~\ref{pic5}. When $\nu=0$, it is a constant and consistent with the $\Lambda$CDM model. The value of $\nu$ has no significant impact on the estimation of the baryon number per comoving volume at the end of recombination if the $\Lambda$CDM model is regarded as a standard (see Table~\ref{table2}).

\section{The effect of running vacuum on the baryon-to-photon ratio}
\label{sec7}

From the previous sections, it is found that running vacuum can indeed influence the particle number densities of photons and baryons by assuming that vacuum is non-minimally coupled to them. In this section, we estimate the possible baryon-to-photon ratio at the end of recombination in RVMs. In cosmology, the baryon-to-photon ratio at different epochs is directly related to our understanding of some processes in the earlier universe. In the $\Lambda$CDM model, the ratio is close to a constant after recombination, and it can even be traced back to the BBN epoch. However, as we saw earlier, in RVMs the particle number densities of photons and baryons could possess different evolution due to running vacuum. Therefore, the baryon-to-photon ratio in RVMs is not a constant even after recombination. For general RVMs, the dynamics term evolves slowly after inflation and the time periods of BBN and recombination are actually very short comparing with the age of the universe, so the influence of running vacuum on BBN and recombination is negligible, i.e., the predicted results of the baryon-to-photon ratio at the ends of BBN and recombination in RVMs are not much different from the results in the $\Lambda$CMD model~\cite{Birkel:1996py,Nakamura:2005qm}. However, the time period from photon decoupling to the present day is almost the age of the universe. The influence of running vacuum on photon number (or baryon number) may be an accumulated effect if there exists a coupling between running vacuum and photons (or baryons)~\cite{Opher:2004vg,Opher:2005px}. Here, we study the influence of such accumulated effect on the evolution of the baryon-to-photon ratio. Since we do not figure out whether running vacuum is more likely to coupled to photons or baryons, we will study both of these two cases.

It is worth noting that when $\nu<0$, the vacuum energy may be negative when the scale factor is small enough [see Figs.~\ref{picex} and \ref{pic5}], which can be also speculated from the definition of the vacuum energy [see Ref.~(\ref{renormalization})]. In general, as we discuss RVMs with $\nu<0$, we must be very careful to insure that the vacuum energy is positive in the early universe. In view of this, to avoid non-physical situations we will only consider the baryon-to-photon ratio in RVMs with $\nu\geq0$.

\begin{figure*}
\begin{center}
\subfigure[]{\label{pic10}
\includegraphics[width=11cm,height=6cm]{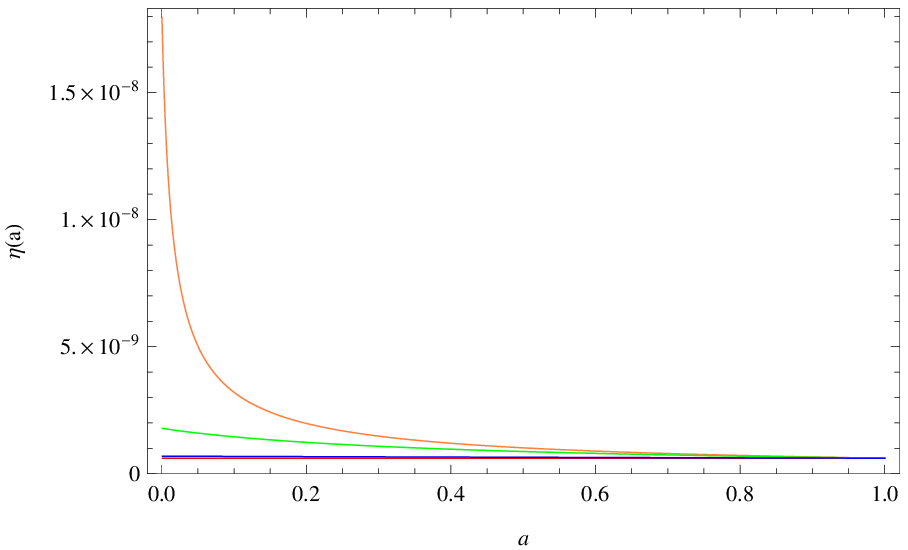}}
\subfigure[]{\label{pic11}
\includegraphics[width=11cm,height=6cm]{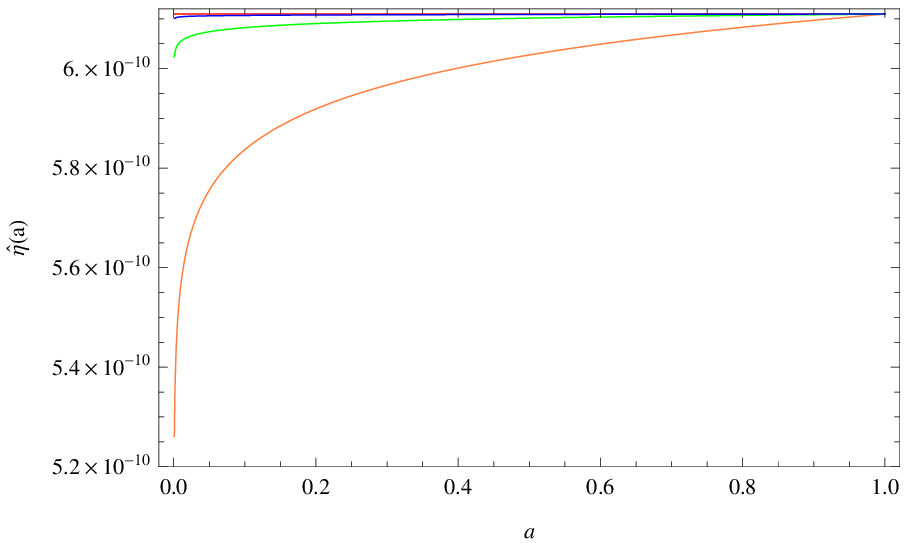}}
\end{center}
\caption{Plots of the baryon-to-photon ratio from recombination to the present day. The upper panel is for the case of running vacuum coupled to photons [$\nu=5.2*10^{-5}$ (orange), $\nu=4*10^{-5}$ (green), $\nu=8*10^{-6}$ (blue), and $\nu=0$ (red)]. The under panel is for running vacuum coupled to baryons [$\nu=10^{-3}$ (orange), $\nu=10^{-4}$ (green), $\nu=10^{-5}$ (blue), and $\nu=0$ (red)]. The current baryon-to-photon ratio is $\eta_0\sim6.11*10^{-10}$.}
\end{figure*}

If running vacuum is only coupled to photons, the range of $\nu$ is about $5.25*10^{-5}>\nu\geq0$, which keeps the energy densities of CMB and running vacuum normal. So, the baryon-to-photon ratio is given by
\begin{eqnarray}\label{00330000}
\eta(a)=\frac{n_b(a)}{n_r(a)}=\frac{n_{b}(a)a^3}{N_{rc}(a)}=
\frac{N_{bc}(a)}{r_A}\left(\frac{T_0^4[a^{4\nu}(1-19024\nu)+19020\nu a]}
{a^4(1-4\nu)}\right)^{-3/4}a^{-3},
\end{eqnarray}
where $N_{rc}(a)$ [see Eq.~(\ref{4200})] is the number of photons per unit comoving volume and $N_{bc}(a)=n_{b}(a)a^3$ is the number of baryons per unit comoving volume, respectively. Since running vacuum is only coupled to photons, the total number of baryons inside the observable universe is a constant after BBN, i.e., $n_{b}(a)a^3=n_{b}(t_0)\sim2.5*10^{-7}$ atoms$\cdot$cm$^{-3}$. From Fig.~\ref{pic10} and Table~\ref{table3}, one can find that when photons are coupled to running vacuum, the baryon-to-photon ratio at the end of recombination, $\eta(a_{re})$, increases with $\nu$, but the deviation is limited by the upper limit of $\nu$. When $\nu=5.2*10^{-5}$, $\eta(a_{re})\sim1.80*10^{-8}$, which is almost 30 times more than the result in the $\Lambda$CDM model (when $\nu=0$, $\eta(a_{re})\sim\eta_0\sim6.11*10^{-10}$). Some other results are also presented in Table~\ref{table3}.

\begin{table}[htbp]
	\centering  
	\caption{The baryon-to-photon ratio at the end of recombination with photons coupled to vacuum}
	\label{table3}  
	\begin{tabular}{|c|c|c|c|c|c|}
		\hline  
		& & & & \\[-6pt]  
		$\nu$ & 0 & $5*10^{-7}$ & $8*10^{-6}$ & $4*10^{-5}$ & $5.2*10^{-5}$\\  
		\hline
		& & & & \\[-6pt]  
		$\eta_{\nu}^{re}$ & $6.11*10^{-10}$ & $6.15*10^{-10}$ & $6.92*10^{-10}$ & $1.79*10^{-9}$ & $1.80*10^{-8}$ \\
		\hline
	\end{tabular}
\end{table}

As we know that $\eta$ can be obtained independently from different epochs of the universe and here are four of the most common epochs: BBN ($\eta^{BBN}$, $z\sim10^9$), recombination ($\eta^{re}$, $z\sim1100$), the period of time associated with the Ly$\alpha$ forest ($\eta^{\text{Ly}\alpha}$, $z\sim2-3$)~\cite{Kirkman:2003uv}, and the present epoch ($\eta_{0}$, $z=0$). If we are clear about the values of $\eta$ at any two epochs of the universe, we can use them to constrain the cosmological models to which $\eta$ is sensitive (see references mentioned above). It also applies to the RVMs in which photons (or baryons) are coupled to running vacuum. Considering the following aspects, it is more reasonable to use $\eta^{re}$ and $\eta_0$ to constrain such RVMs:

1. First of all, $\eta_0$ must be considered because it is given directly by observations, which therefore is the most credible. Moreover, $\eta_0$ is also more accurate than other speculation results: $\eta_0\sim6.11*10^{-10}$~\cite{Aghanim:2018eyx,Tanabashi:2018oca}.

2. Since $\eta_0$ has been chosen, $\eta^{\text{Ly}\alpha}$ will be eliminated. The reason is that these two epochs are too close and so it is hard to inspect the influence of the cosmological model on $\eta$.

3. According to the latest research in the $\Lambda$CDM model, the value of $\eta^{re}$ from the CMB anisotropy is precision: $\eta^{re}\sim(6.090\pm0.060)*10^{-10}$~\cite{Fields:2019pfx}, and the estimation result of $\eta^{BBN}$ from the abundance of light elements has a rather large error: $\eta^{BBN}\sim(6.084\pm0.230)*10^{-10}$~\cite{Fields:2019pfx}. In addition, $\eta^{BBN}$ relies heavily on the expansion rate of the universe. In order to ignore reasonably the influence of running vacuum on the expansion rate of the universe (which is prominent during the very early universe), we employ $\eta^{re}$ rather than $\eta^{BBN}$.

4. The lithium problem in BBN has been unresolved~\cite{Cyburt:2008kw,Fields:2011zzb}, so the estimation of $\eta^{BBN}$ is questionable. There may be new physics beyond the Standard Model at the epoch of BBN.

5. During the period from BBN to recombination, the collision of photons with electrons and atomic nuclei may lead to a change in photon number density. Therefore, the change in $\eta$ after BBN may not only depend on running vacuum.

Reviewing $\eta^{re}$ and $\eta_0$ in the $\Lambda$CDM model, we notice that $\eta_0\sim6.11*10^{-10}$ is almost in the middle of the range of $\eta^{re}\sim(6.090\pm0.060)*10^{-10}$, that is to say, from the end of recombination to the present day, the value of $\eta$ can be reduced or increased on a small scale. We set the current baryon-to-photon ratio in RVMs to equal $\eta_0$ due to the current observations, and $\eta_{\nu}^{re}$ presents the baryon-to-photon ratio in RVMs at the end of recombination. Once again, since we only focus on the cumulative effect of running vacuum on the evolution of the baryon-to-photon ratio after photon decoupling, here we ignore the impact of running vacuum on the CMB anisotropy, that is, we assume that $\eta_{\nu}^{re}$ derived from the CMB anisotropy in RVMs is roughly equal to the estimated value $\eta^{re}\sim(6.090\pm0.060)*10^{-10}$ in the $\Lambda$CDM model. In addition, combining Eq.~(\ref{00330000}) and $\eta_0$, one can also calculate $\eta_{\nu}^{re}$ through the evolution of the number densities of photons and baryons, which is closely related to the evolution of running vacuum. In principle, these two $\eta_{\nu}^{re}$'s should be consistent, so we can make use of the requirement to constrain RVMs. Since the error of $\eta^{re}\sim(6.090\pm0.060)*10^{-10}$ is small, the dynamic term in RVMs must evolve very slowly to ensure that $\eta_{\nu}^{re}$ obtained from Eq.~(\ref{00330000}) in RVMs is not in contradiction with $\eta^{re}\sim(6.090\pm0.060)*10^{-10}$ from the CMB anisotropy.

We list some results of $\eta_{\nu}^{re}$ corresponding to several sets of positive $\nu$ in Table~\ref{table3}. It is seen that when $\nu\geq0$, $\eta_{\nu}^{re}$ is always larger than the current $\eta_0$ and $\eta_{\nu}^{re}$ increases with the parameter $\nu$. We also found that in order not to contradict with $\eta^{re}\sim(6.090\pm0.060)*10^{-10}$, $\nu$ must be at least less than $5*10^{-7}$. In fact, from the perspective of the baryon-to-photon ratio, it is difficult to distinguish RVMs from the $\Lambda$CDM model ($\nu=0$) when $0\leq\nu<5*10^{-7}$. Even if $\nu=8*10^{-6}$, we can find from Fig.~\ref{pic10} that it almost overlaps with the result in the $\Lambda$CDM model. Therefore, if there exits a coupling between photons and running vacuum, according to the data on the baryon-to-photon ratio, the coefficient $\nu$ of the dynamic term of vacuum must be extremely small ($0\leq\nu<5*10^{-7}$). Although we do get a new constraint on $\nu$ by consider the coupling between photons and running vacuum, the new constraint, comparing with $|\nu |<10^{-3}$, is too strong and seems unnatural.

Next, we study the second case, that is, running vacuum is only coupled to baryons, in which the appropriate range of the parameter $\nu$ is $0\leq\nu<10^{-3}$. The baryon-to-photon ratio is
\begin{eqnarray}\label{00330001}
\hat\eta(a)=\frac{N_{bc}(a)}{n_r(a)a^3}=
\left[\left(\frac{317}{49}+\frac{\nu}{245(1+3\nu)}\right)a^{3\nu}
-\frac{268}{49}
-\frac{\nu}{245(1+3\nu)}a^{-1}\right]\frac{n_b(t_0)}{n_r(a)a^3},
\end{eqnarray}
where $N_{bc}(a)$ is given by Eq.~(\ref{0033000}). Note that $\frac{n_b(t_0)}{n_r(a)a^3}=\frac{n_b(t_0)}{n_r(t_0)}
\sim6.11*10^{-10}$ is the current baryon-to-photon ratio for $n_b(t_0)\sim2.5*10^{-7}$ atoms$\cdot$cm$^{-3}$ and $n_r(t_0)\sim4.1*10^{8}\ \text{m}^{-3}$. The evolutionary trend of $\hat\eta(a)$ with the scale factor is shown in Fig.~\ref{pic11}. As the parameter $\nu$ increases, $\hat\eta_{\nu}^{re}$ will be smaller (see Table~\ref{table4}). Comparing these sets of data in Table~\ref{table4}, one finds that for $0\leq\nu<10^{-3}$, $\hat\eta_{\nu}^{re}$ does not change much. Comparing $\hat\eta_{\nu}^{re}$ with $\eta^{re}\sim(6.090\pm0.060)*10^{-10}$, the constraint range of the parameter $\nu$ is reduced to $0\leq\nu<10^{-4}$, which is slightly stronger than $0\leq\nu<10^{-3}$ and seems to be a satisfactory result.

\begin{table}[htbp]
	\centering  
	\caption{The baryon-to-photon ratio at the end of recombination with baryons coupled to vacuum}
	\label{table4}  
	\begin{tabular}{|c|c|c|c|c|}
		\hline  
		& & & & \\[-6pt]  
		$\nu$ & 0 & $10^{-5}$ & $10^{-4}$ & $10^{-3}$ \\  
		\hline
		& & & & \\[-6pt]  
		$\hat\eta_{\nu}^{re}$ & $6.11*10^{-10}$ & $6.10*10^{-10}$ & $6.02*10^{-10}$ & $5.26*10^{-10}$ \\
		\hline
	\end{tabular}
\end{table}

Unfortunately, here the baryon decay rate\footnote{From Fig.~\ref{pic4}, when $\nu\geq0$, the number of baryons per unit comoving volume is monotonically decreasing, which means baryons can only annihilate into vacuum.} is, in fact, rather fast if $\nu$ is not extreme small. One can simply make a following estimation. From recombination to the present day, the average particle number of decaying baryons per year within the observable universe can be estimated as
\begin{eqnarray}\label{003300655}
\bar N=\frac{N_0-N_{re}}{\mathfrak{T}},
\end{eqnarray}
where $N_0$ is the total number of baryons inside the observable universe at present and $N_{re}$ indicates the corresponding value at the end of recombination. Here, $\mathfrak{T}\sim 1.38*10^{10}$ years is the age of the universe (the period before recombination is negligible). If baryon decay is equally likely to occur everywhere in the universe, with such decay rate, the number of baryons decaying into running vacuum on the entire earth per year (for the present day) should be
\begin{eqnarray}\label{00330222}
\bar N_{\text{earth}}=\bar N\frac{N_{\text{earth}}}{N_0},
\end{eqnarray}
where $N_{\text{earth}}$ is the total number of baryons on the earth. In rough, we take the mass of the earth as $6*10^{24}$ kg and the average mass of baryon as $2*10^{-27}$ kg, so $N_{\text{earth}}\sim10^{51}$. The total number of baryons in the current observable universe can be given by the current baryon number density ($n_b(t_0)\sim2.5*10^{-7}$ atoms$\cdot$cm$^{-3}$) and the current volume of the observable universe ($V\sim4*10^{80}$ m$^3$). Given any value of $\nu$, we can obtain a corresponding $N_{re}$ and $\bar N_{\text{earth}}$. Some results are presented in Table~\ref{table5}.

Even if $\nu=10^{-5}$, $\bar N_{\text{earth}}\sim10^{38}\sim10^{-13}N_{\text{earth}}$ is still a huge number, in other words, the earth will lose $10^{11}$ kg due to the decay of baryons into running vacuum each year. Such mass is insignificant compared to the total mass of the earth, but with the current observation accuracy, we may have observed the phenomenon that baryons decay into running vacuum in the laboratory. Obviously, in order not to contradict observations and experiments (such as the half-life of a proton is at least  $1.67*10^{34}$ years~\cite{Bajc:2016qcc}), $\nu$ has to be an extremely small parameter.

\begin{table}[htbp]
	\centering  
	\caption{The number of baryons decaying into running vacuum on the entire earth per year}
	\label{table5}  
	\begin{tabular}{|c|c|c|c|c|}
		\hline  
		& & &  \\[-6pt]  
		$\nu$ &\ \ \  0  \ \ \ & $10^{-5}$ & $10^{-4}$ & $10^{-3}$ \\  
		\hline
		& & &  \\[-6pt]  
		$\bar N_{\text{earth}}$ & 0 & $\sim10^{38}$ & $\sim10^{39}$ & $\sim10^{40}$ \\
		\hline
	\end{tabular}
\end{table}

\section{Discussions and conclusions}
\label{sec8}
The energy-momentum tensor conservation of photons is not compatible with the energy conservation of photons in the $\Lambda$CDM model, and the latter is usually not satisfied because of the expansion of space-time. In RVMs, it is found that if $\rho_{\Lambda}=\frac{3}{4\kappa^2} H^2+c_0$, the energy conservation of photons can be satisfied in the context of a toy cosmological model. And if $c_0=0$, the total energy of the universe can be conserved. However, these conditions can not be implemented for a realistic universe due to $\rho_b(t_0)+\rho_{dm}(t_0)\neq0$. Therefore, for a realistic universe, the energy of photons (and also the universe) is not conserved even in the context of RVMs.

The value of the baryon-to-photon ratio, which is approximately a constant in the $\Lambda$CDM model, at different epochs indicates the characteristics of the corresponding cosmological processes. One can estimate the baryon-to-photon ratio at various epochs based on the cosmological processes, which is a crucial tool to check the availability of cosmological models. The study on the baryon-to-photon ratio in the standard $\Lambda$CDM model explains the current abundance of most light elements in the universe based on BBN~\cite{Malaney:1993ah,Sarkar:1995dd,
Tytler:2000qf,Fields:2014uja,Mathews:2017xht,Zyla:2020zbs}, which is an important basis for BBN to be recognized. However, the difference between the estimation and observation on lithium abundance is like a dark cloud hanging over BBN and the $\Lambda$CDM model~\cite{Cyburt:2008kw,Fields:2011zzb}. Although varying physical ``constants'' could bring about a turning point to the problem of lithium abundance~\cite{Gupta:2020wgz,Gupta:2021ioh}, we do not know if varying physical ``constants'' will bring about new problems. Therefore, we study the baryon-to-photon ratio in RVMs.

Since there exists a non-minimal coupling between matter fields and running vacuum in RVMs, the density of certain substance will be affected by running vacuum. If the substance is photons or baryons, the baryon-to-photon ratio must evolve with running vacuum. Considering that the photon density may be affected by a variety of factors before the end of recombination, in order to analyze and study the influence of running vacuum on the baryon-to-photon ratio without other distractions, we study the period from the end of recombination to the current universe. For photons coupling to running vacuum, the temperature of CMB at the end of recombination, still needs to meet the requirement of photon decoupling, and it is almost independent on the model of the universe. According to the current temperature of CMB, we can inversely deduce the size of the scale factor at recombination, which changes with the coefficient $\nu$ in running vacuum (see Table~\ref{table1}). The photon number of the observable universe at the end of recombination is also related to $\nu$ [see Fig.~\ref{pic2}]. Similarly, we obtain the baryon density [see Fig.~\ref{pic3}] and baryon number density [see Fig.~\ref{pic5}] for different values of $\nu$, when running vacuum is coupled to baryons. In order to ensure that the vacuum energy is positive in the early universe, the parameter $\nu$ must be equal or greater than 0. When photons are coupled to running vacuum, the upper limit of $\nu$ is approximately equal to $5.25*10^{-5}$, which does not appear for the coupling between baryons and vacuum.

Combining the current observations of the baryon-to-photon ratio and the corresponding value at recombination predicted by the CMB anisotropy, we can get new constraints on RVMs by considering that running vacuum is coupled to photons or baryons. When photons are coupled to running vacuum, to guarantee that the baryon-to-photon ratio in RVMs does not conflict with the CMB anisotropy, the parameter $\nu$ must be abnormally small. Such a new constraint on $\nu$ is unnatural. When baryons are coupled to vacuum, we find that the value of $\nu$ just needs to be $0\leq\nu<10^{-4}$ to satisfy the same requirement. However, the decay rate of baryons into vacuum is too fast when $\nu\to10^{-4}$. Our estimations show that unless $\nu$ is extremely small, we should be able to observe the decay process of non-conservation of baryon number on the earth.

In summary, it can be concluded that, in RVMs the possibility of running vacuum coupling to photons or baryons is almost zero. At present, we do not figure out whether the running vacuum can be coupled to dark matter and how to determine the possibility of such a coupling. It is a question worthy of further discussion. Our research on constraining RVMs is also applicable to other cosmological models in which the baryon-to-photon ratio may evolve over time.

\section*{Acknowledgments} \hspace{5mm}
This work was supported by the National Natural Science Foundation of China under Grant Nos.~11873001 and 12005174. K. Yang acknowledges the support of Natural Science Foundation of Chongqing, China under Grant No. cstc2020jcyj-msxmX0370. J. Li acknowledges the support of Natural Science Foundation of Chongqing (Grant No. cstc2018jcyjAX0767).

\providecommand{\href}[2]{#2}\begingroup\raggedright

\end{document}